\documentclass[twocolumn,showpacs,aps,prb,a4]{revtex4}
\usepackage{dcolumn} 
\usepackage{graphicx}
\usepackage{subfigure}
\usepackage{epsfig}
\usepackage{epstopdf}
\begin{document}
\title{Fermi-surface pockets in YBa$_2$Cu$_3$O$_{6.5}$ : A comparison of \textit{ab initio} techniques}
\author{Danilo Puggioni, Alessio Filippetti, and Vincenzo Fiorentini}	

\affiliation{CNR-INFM SLACS and Dipartimento di Fisica, Universit\`a di Cagliari,
Cittadella  Universitaria, I-09042 Monserrato (CA), Italy\\}

\date{}

\begin{abstract} 

We study the Fermi surface of  metallic, non-magnetic \textit{ortho}-II YBa$_2$Cu$_3$O$_{6.5}$ using three different 
density-functional-based band-structure  techniques (GGA, GGA+U, PSIC). The calculated Fermi surface exhibits no pockets in GGA+U and PSIC, a minor one in GGA. Upon shifting the Fermi level in the vicinity of the calculated value, we instead observe several pocket structures. We
calculate their cross-sectional areas and cyclotron masses. Overall, our calculations show no solid evidence of the existence of electron-like --nor, in fact, of any-- Fermi surface pockets in this phase. This suggests that the origin of the pockets should be sought for in other, different phases.

\end{abstract} 
\maketitle

\section{Introduction}
The Fermi surface  of underdoped high-temperature cuprate superconductors  is currently under intense  investigation. 
Recently \cite{doiron,leboeuf,jaudet,twopockets} Shubnikov-de Haas (SdH) and de Haas-van Alphen (dHvA) oscillations were observed in \textit{ortho}-II YBa$_2$Cu$_3$O$_{6.5}$ (henceforth YBCO; \textit{ortho}-II stands for the chain-aligned oxygen configuration with one Cu(1)-O 
chain per 2$\times$1$\times$1 cell). These oscillations (of resistance and Hall coefficient in SdH, and magnetization in dHvA) correspond to closed sections (pockets) of the Fermi surface and they exhibit, as a function of the inverse of the magnetic field, characteristic frequencies related to the cross-sectional area of the pocket (or pockets: their number and location is undetermined). 

The frequency measured by dHvA experiments (more accurate than SdH) is 540$\pm$4 T, corresponding to a small portion (2\%) of the Brillouin zone being enclosed by the  pockets. The cyclotron mass, deduced from a Lifshitz-Koshevic fit of  the oscillation amplitude vs temperature, is m=1.76$\pm$0.07 free-electron masses. The oscillations were observed\cite{doiron,leboeuf,jaudet} in high field and only 
at low (4 K) temperature; the sign of the Hall coefficient was seen to become negative from about 25 K downward, and this was interpreted as
a signature of the pockets in question being electron-like in nature. A further recent measurement\cite{twopockets} in YBa$_2$Cu$_3$O$_{6.51}$ reported, in addition to the same signal of Ref.\onlinecite{doiron}, an oscillation  with frequency and mass in the vicinity of 1600 T and 3.4 m$_e$ respectively, allegedly (see Ref.\onlinecite{twopockets}, p.201) associated with a hole-like pocket.

SdH and dHvA examine at low temperature a  state obtained by applying a high magnetic field to the superconductor.  To a first approximation this state is supposed to be the normal (possibly pseudo-gap) state. The simplest  hypothesis is that once superconductivity is removed, YBCO is a metallic and non-magnetic Fermi-liquid like system (although more sophisticated options also exist, such as e.g. magnetic fluctuations and polaronic formations in stripe-like morphology\cite{lake} and more).  Since experiments are often interpreted based on this assumption, an  issue to be settled is whether or not the Fermi surface of this specific non-magnetic metallic phase  exhibits pockets as revealed in experiments.  If no calculated pockets exist, or can be identified with those observed, then some other phase will have to be invoked as the state accessed in oscillation experiments. To address this issue, here we employ three distinct techniques based on density-functional theory (DFT): GGA (generalized gradient approximation), GGA+U, and PSIC (pseudo-Self-Interaction-Correction) method. Furthermore, we adopt the common practice  (discussed below) of applying rigid-band shifts to explore the Fermi surface in a wide energy interval surrounding the calculated E$_{\rm F}$. Our calculation widen the scope of recent\cite{esempio}  calculations limited to the GGA approach.

Our  study shows that overall there is no reliable indication  that non-magnetic metallic YBCO possesses electron-Fermi surface pockets. Specifically, only one technique (the GGA+U) finds an electron-like pocket, appearing however at a --60 meV shift away from calculated E$_{\rm F}$. None of the other techniques find any such pocket in a $\pm$100 meV interval around E$_{\rm F}$. As we will argue, in fact, there is only scant evidence for hole-like pockets as well. 

While we do not question the reliability of the SdH/dHvA experiments, we note that one may envisage ways to generate a negative Hall coefficient other than the  existence of  electron-like pockets. For example, oscillations may be due to hole-like pockets, and the Hall coefficient positive-to-negative crossover may stem from the contribution of other electron-like Fermi surface structures, mixed up by differently temperature-dependent electron and  hole mobilities. Other considerations that must be marked  on the theoretical roadmap are that the pocket structure is partially at odds with the ``Fermi arcs'' observed \cite{hoss, norman, shen,schabel, lu, zabo} in angle-resolved photo-emission spectroscopy; and that ordering phenomena, possibly related to magnetic structure or density waves, may be causing a reconstruction of the Fermi surface.

\section{Method}

We calculate the band structure of YBCO in the non-magnetic metallic state with three different  DFT-based techniques. We assumed the crystal structure of YBa$_2$Cu$_3$O$_{6.5}$ determined by Grybos \textit{et al}.\cite{grybos1, grybos2}   We use the GGA 
(generalized gradient approximation), GGA+U, and  the pseudo-self-interaction correction method (PSIC), a parameter-free, first-principles 
DFT-based method\cite{Filippetti1} which correctly describes the physics of several correlated cuprates,\cite{Filippetti2,Filippetti5,Filippetti6} and 
yet is practically viable for large-sized systems. In particular the PSIC  is able to describe the competition of metallic and insulating phases of  YBa$_2$Cu$_3$O$_{6+x}$ from $x$=0 (where it is\cite{Filippetti6} an antiferromagnetic Mott 
insulator) across two metal-insulator transitions\cite{Filippetti5} to metallic $x$=1, obviating to the failures of plain GGA or similar approaches in this context.

Our  GGA and GGA+U calculations are carried out using the VASP package\cite{kresse1,perdew} with the projector-augmented wave method 
(PAW).\cite{blochl} The PSIC calculation are performed using a custom in-house code  with ultrasoft pseudopotentials\cite{uspp} and a plane wave 
basis set. The cutoff energy was set at 420 eV. A Monkhorst-Pack\cite{mp} 9$\times$19$\times$6 grid was used for the self-consistency cycle. 
We intentionally used the in-plane 2$\times$1 periodicity appropriate to chain-ordered \textit{ortho}-II YBCO at this specific doping, since experiments are claimed to be performed in this structure.  We tested non-spin-polarized calculations, spin-polarized calculations with small initial moments, and fixed-magnetic-moment calculations with zero imposed magnetization, consistently getting the same results, i.e. a non-magnetic metallic state. The Fermi surfaces are visualized with the Xcrysden  package.\cite{xcrysden} 

We used the Dudarev implementation\cite{dudarev} of GGA+U, whereby the relevant parameter is the difference $U$--$J$ of the effective
on-site Coulomb  and exchange interactions. $U$--$J$ was set to 9 eV for the $d$ states of planar Cu, the value reproduces the fundamental gap of Mott-insulating antiferromagnetic YBa$_2$Cu$_3$O$_6$ as obtained in PSIC\cite{Filippetti5,Filippetti6} or in  experiment (no qualitative changes are observed down to $U$--$J$=6 eV for YBCO). We underline that the paramagnetic Fermi surface calculation is sensitive to $U$--$J$ via small orbital polarizations (i.e. deviations from exact half-filling) in the partially occupied Cu $d$$_{x^2-y^2}$ states, and this may affect the details of band morphology in the vicinity of E$_{\rm F}$. 

\section{Results and discussion}

\subsection{Band structures}

\begin{figure}[h]
\includegraphics[clip,width=8.8cm]{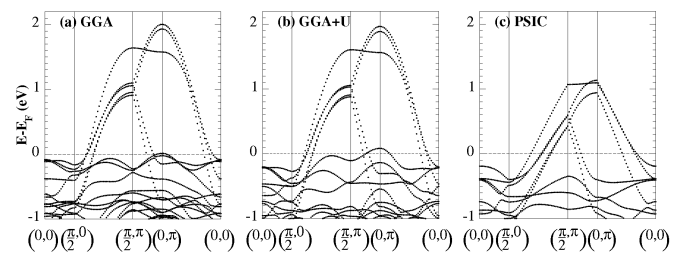}
\caption{Band structures of YBa$_2$Cu$_3$O$_{6.5}$. (a) GGA, (b) GGA+U with $U$--$J$= 9 eV, (c) PSIC.}
\label{band}
\end{figure}

In Fig.\ref{band} we compare the band structures in the k$_z$=0 plane, as obtained by the three methods. The dispersion in k$_z$ is weak and not 
important in the present context. k$_x$ and k$_y$ are in units of the inverse 1/$a$ and 1/$b$ of the in-plane lattice constants. The leftmost panel
(Fig.\ref{band}(a)) displays the bands within the GGA approach. Moving along the ($\pi$/2,0)-($\pi$/2,$\pi$) direction, the first band to cross 
E$_{\rm F}$ is mainly due to states of the Cu(1)-O chain. This band is very close to being one dimensional. The next four bands crossings E$_{\rm F}$ come from the CuO$_2$ planes.
There is a splitting between the bonding and antibonding CuO$_2$ bands of $\sim$0.2 eV along the ($\pi$/2,0)-($\pi$/2,$\pi$) line at E$_{\rm F}$. 
Each of these two bands are further split up by the additional 2$a$ periodicity (this is most evident close to the point ($\pi$/2,$\pi$)). 
We find that the splitting is 40 meV at E$_{\rm F}$ along the ($\pi$/2,0)-($\pi$/2,$\pi$) direction. 
In the GGA calculation a fairly flat Cu(1)O chain-O$_{\rm apical}$ band crosses E$_{\rm F}$  close to the (0,$\pi$) point and gives rise to a small
tubular quasi-2D hole pocket. This band is 13 meV above E$_{\rm F}$ at (0,$\pi$). In addition, a second band with a similar character lies 
just 20 meV below E$_{\rm F}$ at (0,$\pi$). Our results are similar to calculations on YBCO reported previously.\cite{esempio,bascones}

In the central panel, Fig.\ref{band}(b), we show the GGA+U bands. Overall, the GGA+U rendition appears quite close to those of GGA. This is
expectable as U only affect magnetic and/or orbital-polarized states, thus the paramegnetic configuration is mildly affected.   
The main difference with respect to the GGA case is that the flat chain-apical bands crossing E$_{\rm F}$ near (0,$\pi$) are now about 80 meV 
above E$_{\rm F}$ and 140 meV below E$_{\rm F}$ at (0,$\pi$), i.e. they are split by more than 200 meV, compared to about 30 meV in GGA. 
This difference is due to the indirect (i.e. self-consistent) effect of the orbital polarization of in-plane Cu d states on the band manifold.\cite{dudarev}

The right panel, Fig.\ref{band}(c), shows our calculation with the PSIC technique. Here we see more radical differences with respect to 
the other two methods, mainly due to the fact that PSIC corrects for sef-interaction Cu $d$ as well as O $p$ state occupations, so that  the corrections can be equally sizable for non magnetic and/or non orbitally-polarized states. This description results in generally less dispersed
band structure; chain bands are now far from E$_{\rm F}$, and the net result is that there are no small pockets in the Fermi surface.

\subsection{Fermi surfaces}

Strictly speaking, the theoretical prediction of the Fermi surface is based on the calculated electronic structure and Fermi level.  Here, however, we  also consider how the Fermi surface changes upon an upward or downward shift of the Fermi level compared to the calculated value.
This is a fairly common practice in band theory studies of superconductors. The first motivation is that, while  DFT calculations usually describe well the general features of the band structure of metals, small discrepancies
in the relative positions of the bands are common when comparison with experiment is involved. (Generally, this relates to structural details and of course  to the DFT description of the electron correlation.) 
For example, in Sr$_2$RuO$_4$, studied in detail with the dHvA technique, the Fermi energy needs to be shifted by 40 meV in either 
direction\cite{mackenzie} to improve the calculated-bands agreement  with experiment. Even in MgB$_2$, shifts of the order of 100 meV are 
needed.\cite{carrington1, carrington2} 

A further motivation pertaining to doped cuprates is that Fermi level shifts roughly simulate doping fluctuations. Of course the shift-doping relation  depends on which specific band or bands are or get occupied upon shifting. In our case the maximum shifts applied ($\sim$50--60 meV) correspond to rather substantial  doping fluctuation ($\sim$$\pm$0.04, i.e. a 30\% of the nominal doping).

\begin{figure}
\includegraphics[clip,width=8.9cm]{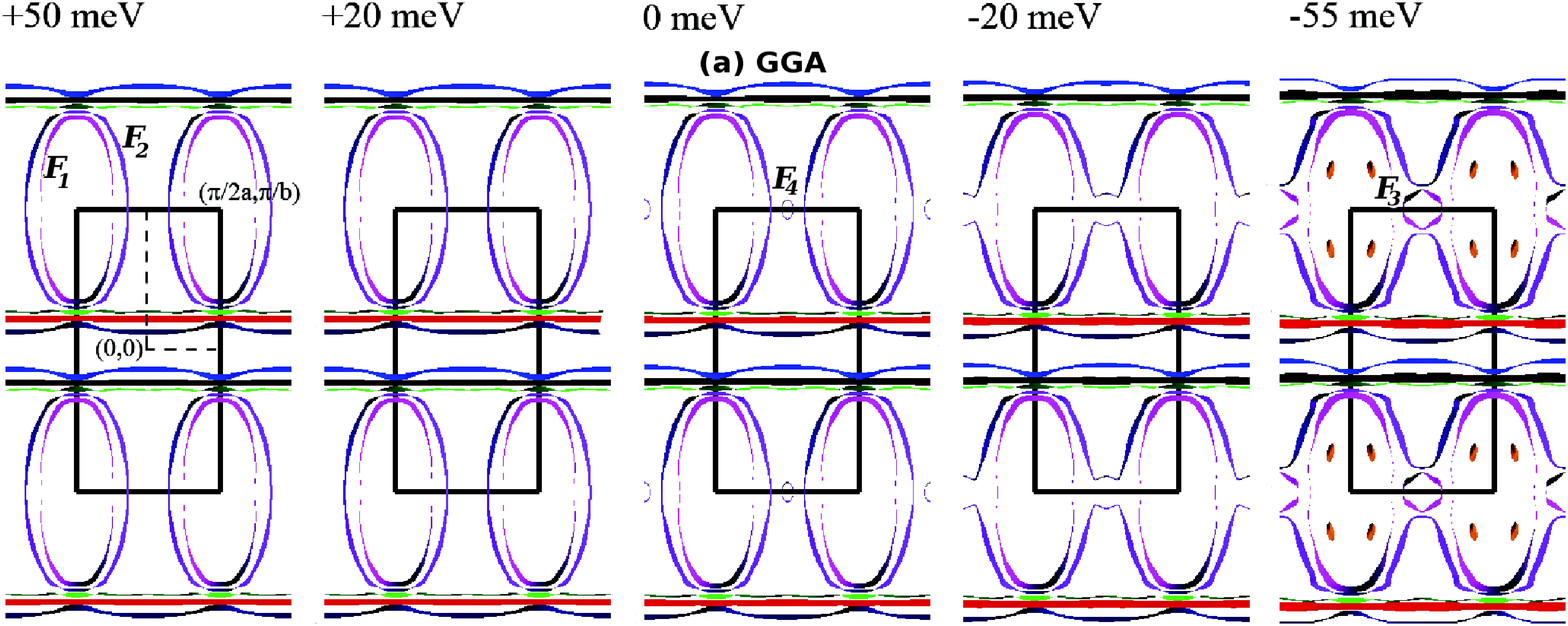}\\
\includegraphics[clip,width=8.9cm]{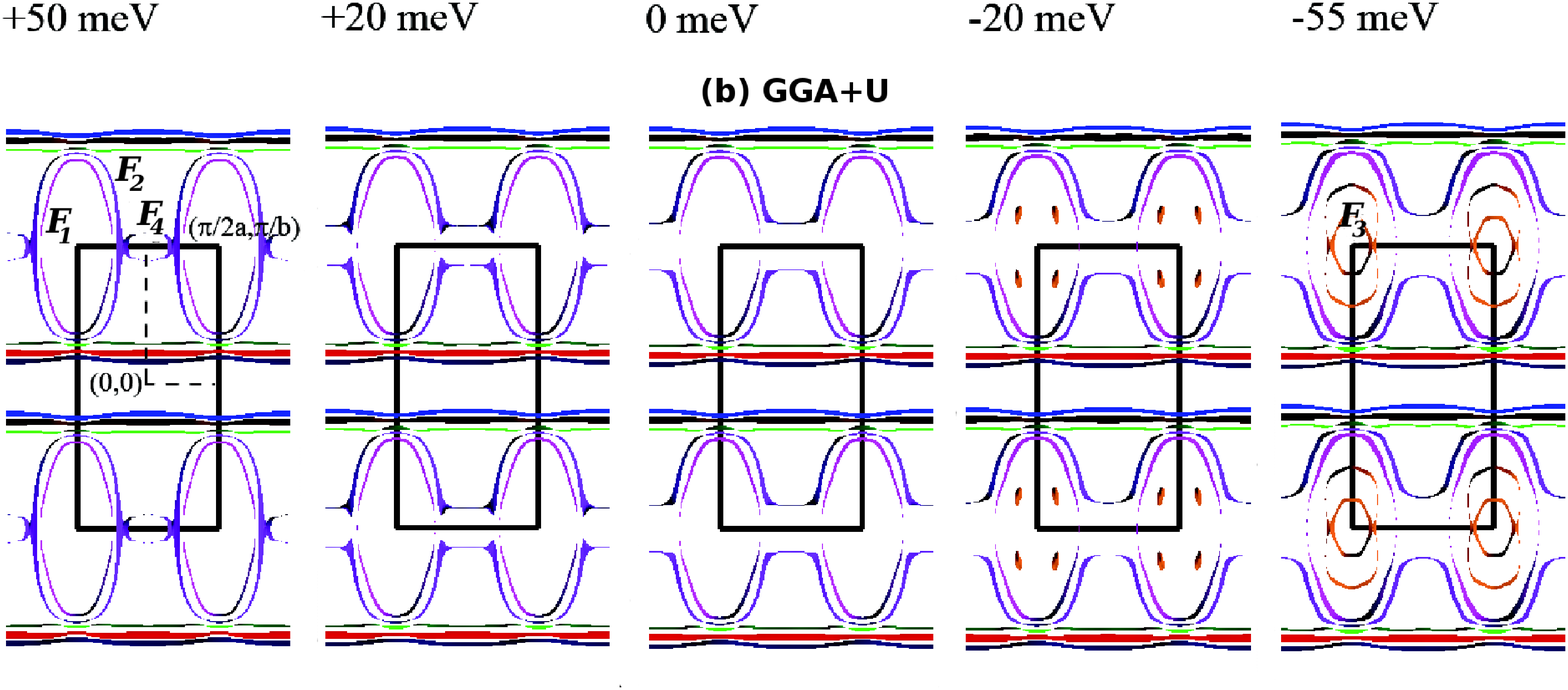}\\
\includegraphics[clip,width=8.9cm]{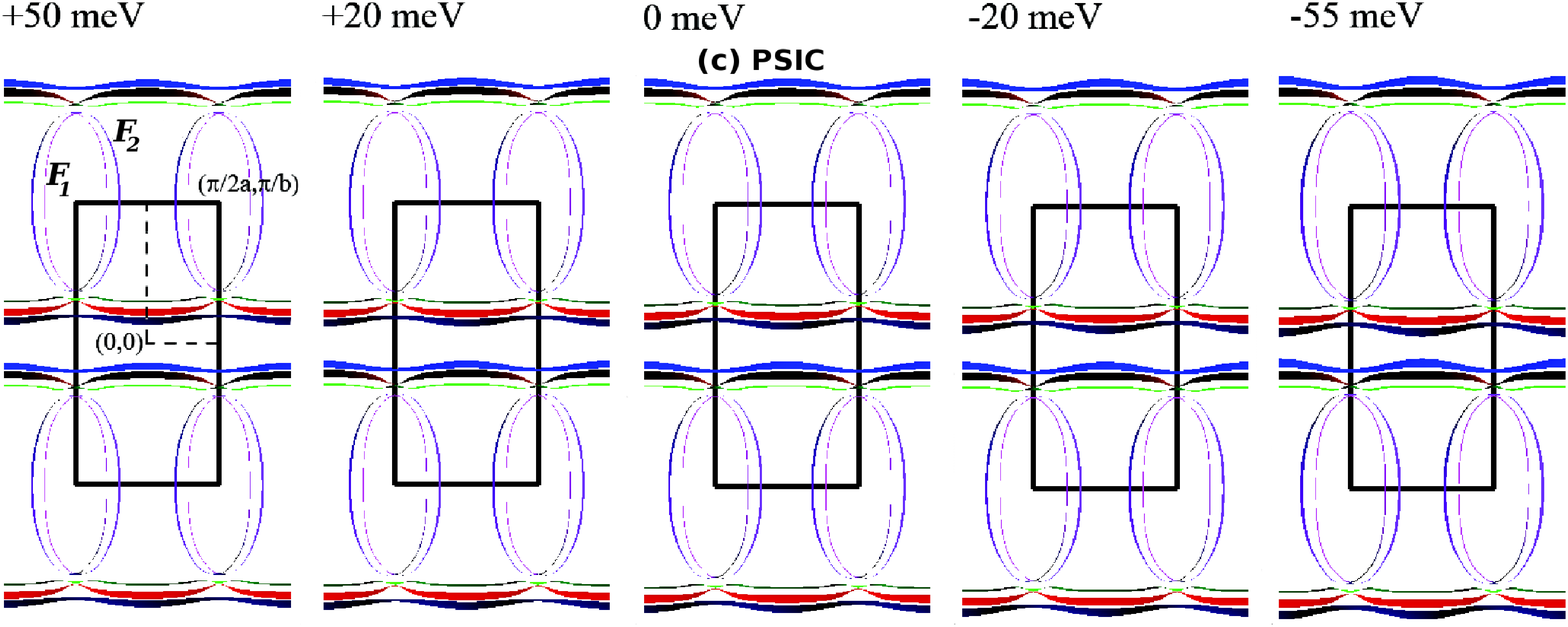}\\
\caption{Color on-line) Evolution of the Fermi surface of YBCO with Fermi-level shift $\Delta$E$_{\rm F}$ in the basal plane ($k_z=0$). 
The main quantum oscillation orbits ($F_n$) are marked on the O meV and --55 meV panel for GGA (a), on the +50 meV and --55 meV for GGA+U (b). In PSIC  there are no small  pockets.}
\label{fermi}
\end{figure}

In Fig.\ref{fermi} we collect the Fermi surface for the three techniques (top to bottom), and upward to downward (left to right) shifts of 
the Fermi level. The top panel (Fig.\ref{fermi}(a))  reports GGA results. For $\Delta$E$_{\rm F}$=+50 meV the Fermi surface consists  of just
two large hole-like CuO$_2$ sheets centered on ($\pi$/2,$\pi$), plus three quasi-one-dimensional sheets (one from the chains, and two from the 
planes). As E$_{\rm F}$ shifts down, a small hole-like pocket develops near the (0,$\pi$) point, originating from the flat CuO-O$_{\rm apical}$
band discussed earlier. A further lowering of E$_{\rm F}$  causes this pocket to grow in size and then merge with the antibonding CuO$_2$  plane
sheet. As E$_{\rm F}$ is further reduced, the second CuO-BaO band crosses the Fermi level, giving rise to another pocket. Eventually, this merges
with the bonding CuO$_2$ plane sheet. Similar results were recently reported in Refs.\onlinecite{esempio} and \onlinecite{elfimov}.

Fig.\ref{fermi}(b)  shows the Fermi surface evolution according to  GGA+U calculations. In this case for $\Delta$E$_{\rm F}$=+50 meV the Fermi surface is similar 
to the GGA calculation, but shows a hole-like pocket near the (0,$\pi$) point, whose origin is the chain-apical band. 
This pocket merges with the CuO$_2$ sheets at zero shift. This trend is again expected given the larger splitting of the chain-apical band 
at (0,$\pi$) discussed in connection with Fig.\ref{band}. For $\Delta$E$_{\rm F}$=--55 meV, an electron-like pocket appears near ($\pi$/2,$\pi$),
surrounded by a hole-like sheet. Going back to Fig.\ref{band}, one immediately realizes that this is also due to the enhanced splitting in GGA+U: a similar pocket would appear in GGA  for a much larger  negative shift of over 200 meV.

Fig.\ref{fermi}(c) shows the PSIC results. The only  structures in the Fermi surface are two large hole-like CuO$_2$ sheets
centered on ($\pi$/2,$\pi$). The Fermi level shift  only moderately affect their area.  No small pockets appear in this shift interval.

Overall Fig.\ref{fermi} shows a marked sensitivity of the GGA and GGA+U Fermi surface to the relative positions of the bands. This suggests that subtle changes
in doping could result in the formation of small Fermi surface pockets. On the other hand, the PSIC Fermi surface is quite independent of doping, and would lead to predict or expect no small pockets at all. 

\subsection{Fermi surface pockets: frequencies and masses}

\begin{figure}
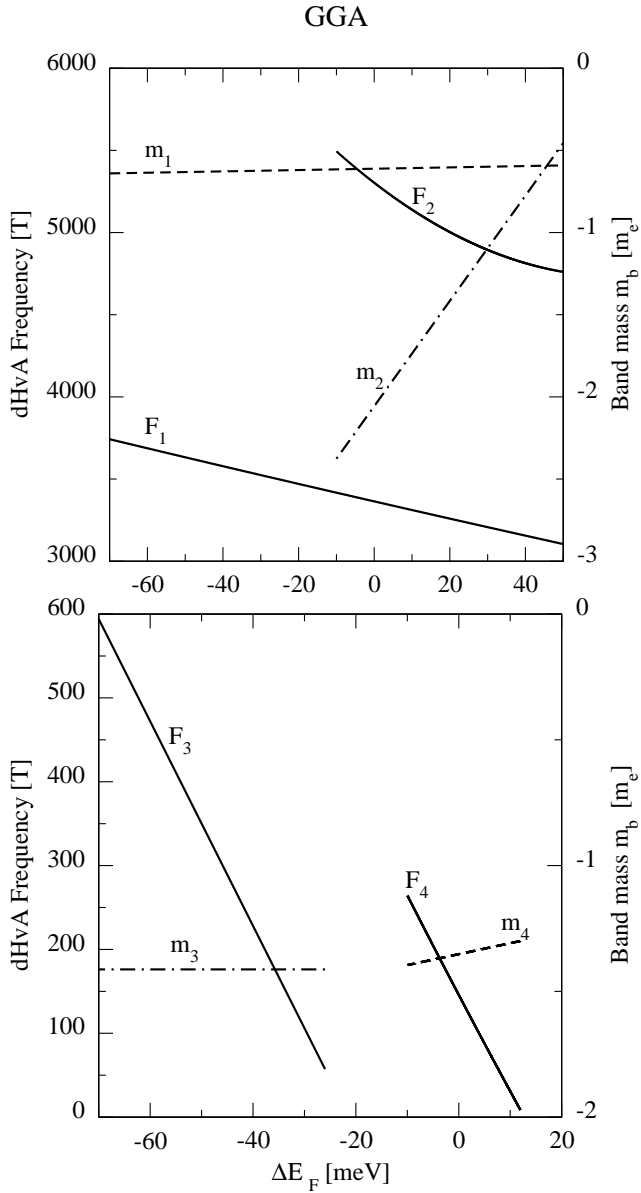

\includegraphics[clip,width=8.4cm]{fig3a.eps}\\
\includegraphics[clip,width=8.4cm]{fig3b.eps}
\caption{GGA extremal dHvA frequencies (solid) for the large hole-like pockets (a) and for the small pockets (b). Dash-dotted lines 
are the calculated band masses.}\label{extremalgga}
\end{figure}

\begin{figure}
\includegraphics[clip,width=8.8cm]{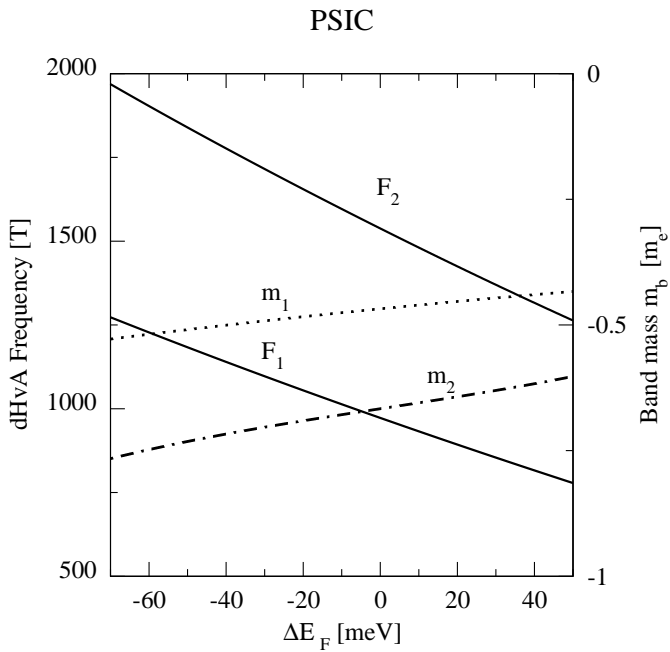}
\caption{PSIC  extremal dHvA frequencies (solid) for the large hole-like pockets. Dash-dotted lines are the calculated band masses.} 
\label{extremalpsic}
\end{figure}

To make contact with the quantum oscillations  measured in SdH and dHvA experiments,\cite{doiron,leboeuf,jaudet,twopockets} we calculate the quantum oscillation frequencies F=($\hbar$$A$/2$\pi e$) from the cross-sectional area $A$ of the orbits (i.e. the pockets), and the attendant cyclotron 
masses m=$\hbar^2$(${\partial A}$/${\partial E}$)/2$\pi$ for the various structures found by the different techniques. They are reported in Figs. \ref{extremalgga}, \ref{extremalggau}, and \ref{extremalpsic}
for GGA, GGA+U, and PSIC, respectively. 

\begin{figure}
\includegraphics[clip,width=8.4cm]{fig4a.eps}
\includegraphics[clip,width=8.4cm]{fig4b.eps}
\caption{GGA+U  extremal dHvA frequencies (solid) for the large hole-like pockets (a) and for the small pockets (b). Dash-dotted lines 
are the calculated band masses.}\label{extremalggau}
\end{figure}

For all techniques  we report  the  high-frequency oscillations  related to large cylinders; for GGA and GGA+U only, low-frequency oscillations
related to small pockets are reported in a second panel. Thus, the frequencies shown in Fig.\ref{extremalgga}(a), Fig.\ref{extremalggau}(a) and 
Fig.\ref{extremalpsic} (F$_1$  and F$_2$) are from the main CuO$_2$ sheet surfaces, whereas those in Fig.\ref{extremalgga}(b) and Fig.\ref{extremalggau}
(b) (F$_3$ and F$_4$) are from the small pockets. We note, first of all, that the frequencies calculated for the main CuO$_2$ sheets (F$_1$  and F$_2$)
are similar for GGA and GGA+U with frequencies between 3000 T and 5500 T, whereas the frequencies calculated with the PSIC approach are between 
1000 T and 2000 T. The reason of this difference is the lesser dispersion of the band structure as calculated with the PSIC technique. 
All values are way larger than the experimental one; the masses are typically  a factor of two (or more) smaller than in experiment, and always negative. These Fermi surface sheets can therefore be ruled out as the origin of the experimental oscillations reported so far.

Next we analyze the small-pocket signals in the frequency range 0 to 900 T. In the GGA calculation, the hole-like pocket F$_3$ has a frequency
between 100 and 600 T depending on the E$_{\rm F}$ shift; the experimental value would be attained at a shift of about --65 meV. The calculated mass
of this pocket  is  shift-independent, and equal to $\sim$--1.4 m$_e$. The F$_4$ pocket has a fairly low frequency of 100 to 300 T and a negative
mass similar to F$_3$.  With the GGA+U approach we find the hole-like pocket F$_4$, with a roughly shift-independent  mass of $\sim$--1.25 m$_e$ and frequency in the 400 to 600 T range, and the electron-like pocket F$_3$ with frequency between 400 and 800 T and a sharply
varying mass, between 1.5 m$_e$ and 2.3 m$_e$. 

Comparing with experiments,\cite{doiron,leboeuf,jaudet} several of our calculated pockets may seem good candidates. Frequencies and  masses (in absolute value) are more or less in the ballpark. If we accept the assumption that the change of sign of the Hall resistance\cite{leboeuf} is purely due to the electron-like nature of the pockets, we implictly fix the experimental sign of the mass to a positive value. The  frequency {\it and} mass deduced from  observation\cite{doiron,leboeuf} would then be compatible only with the F$_3$ GGA+U pocket.
 
A very recent  measurement\cite{twopockets} in YBa$_2$Cu$_3$O$_{6.51}$ has revealed, in addition to the same signal of Ref.\onlinecite{doiron}, an oscillation with frequency and mass in the vicinity of 1600 T and 3.4 m$_e$ respectively.  In Ref.\onlinecite{twopockets} (p.210) the signal is attributed  tentatively to a hole-like pocket. In all our calculations, including shifts, there is only one case (GGA+U at large negative shift, rightmost picture in panel b of Fig. \ref{fermi}) in which  hole and electron pockets coexist. Near the ({$\pi$/2,$\pi$) point in the GGA+U calculations, starting at a shift of --55 meV, the structure recognizably involves two distinct pockets: one is the electron pocket F$_3$ discussed above; the other a larger hole-like pocket surrounding F$_3$ itself. Their simultaneous presence is due to a change in curvature of the same band, most notably between ({$\pi$/2,$\pi$) and ({$\pi$/2,0). The character of this band is, like that of F$_3$, strongly chain-apical. The corresponding calculated frequency is about 
2200 T and a mass of --1.4 m$_e$.   The frequency is very  roughly similar to the 1600 T measured in Ref.\onlinecite{twopockets}, while the mass is over a factor two smaller.

Overall, however, we conclude that there is not enough evidence to actually associate our calculated results to the experimental findings of Refs. \onlinecite{doiron,leboeuf,jaudet,twopockets}.  The reasons will be discussed in the next Section.

\subsection{Discussion and summary}

The calculations just reported have detected several small pockets (mainly hole-like) roughly compatible with the observed 
oscillation. However, all these small pockets have essentially chain or chain-apical character, and not in-plane character.
GGA+U does seemingly finds the ``right" pattern of coexisting electron and hole pockets, but (aside from the need for an artificial --60-meV Fermi level shift, corresponding to a 30\% overdoping) both pockets have
a chain-apical nature even stronger than the corresponding GGA-calculated band due to the remarkable (perhaps exaggerated) $U$-induced lowering
of in-plane Cu bonding states.   

On the other hand, there appears to be experimental evidence that the negative and oscillating Hall resistance at low temperature resulting
from  electron-like pockets (i.e. a positive mass)  be related to states residing in the CuO$_2$ planes.  This is supported\cite{doiron}
by the suppression of $ab$-plane conductivity anisotropy below 100 K, implying that chains  do not conduct  at low temperatures (and high field). 

Further supporting the fact that Fermi surface pockets are a plane-related feature, quantum oscillations were observed in
YBa$_2$Cu$_4$O$_8$.\cite{leboeuf,yelland, bangura} Calculations\cite{bangura, draxl, esempio} for that compound have shown that  the GGA-calculated band related to the F$_3$ hole pocket in YBa$_2$Cu$_3$O$_{6.5}$ is now as far as 400 meV below E$_{\rm F}$, hence cannot not  reasonably invoked to explain the observations. Consistently, we found (unpublished calculations)
that no pockets appear at all in the non-magnetic phase of chainless Y$_{0.75}$Ca$_{0.25}$Ba$_2$Cu$_3$O$_6$.

We further recall that pockets appear only upon appreciably shifting the Fermi energy: the proper calculated Fermi surfaces, i.e. those at zero shift, show no small  pockets, except for the GGA F$_4$ hole pocket 
of Fig. \ref{fermi}(a), related to  the backfolding in  the 2$\times$1 cell of a pocket found by GGA itself in YBa$_2$Cu$_3$O$_{7}$ (not seen by  ARPES).

Were we forced to embrace one of the  methods applied here and the pertaining conclusions as the most reliable in this context, we would by all means pick PSIC, and conclude that in non-magnetic YBCO simply there are no small pockets, electron-like or otherwise. Indeed, among those used here, PSIC has shown to be by far the most dependable technique in the context of cuprates. For instance, the energy balance of various magnetic phases of  YBa$_2$Cu$_3$O$_{6+x}$ is correctly described, and so are the general properties of a number of cuprates.\cite{Filippetti2,Filippetti5,Filippetti6}  Furthermore, in the context of Fermi surface determination, PSIC matches ARPES perfectly for YBa$_2$Cu$_3$O$_{7}$ (unpublished calculations) whereas GGA finds, as mentioned, a  zone-corner pocket which ARPES does not observe.

In summary, we  presented calculations of the electronic structure of YBCO in the non-magnetic state with three different DFT-based approaches: GGA, GGA+U and PSIC. Upon substantial shifts of the Fermi energy, GGA and GGA+U do produce small Fermi surface pockets, mostly originating from chain or chain-apical bands, with frequencies and band masses similar to those experimentally observed (one GGA+U pocket has a positive cyclotron mass, i.e. is electron-like), while PSIC shows no small pocket at all.  As discussed, our conclusion is that there is no unambiguous evidence for the existence of electron-like pockets --nor, indeed, of any pockets-- in the non-magnetic metallic state of YBa$_2$Cu$_3$O$_{6.5}$.  In addition, no pockets (either electron or hole) derive from in-plane states. This is a conclusion coherently obtained by three different \textit{ab initio} techniques. We  suggest that the experimentally observed pockets are a property of {\it another} state of YBCO, possibly characterized by some form of ordering (probably magnetic, given its  coexistence with superconductivity up to high doping revealed by many experiments) causing a Fermi surface reconstruction. We will present elsewhere further first-principles work  in this direction.

\section*{Acknowledgments}

Work supported in part by MiUR through PRIN 2005 and  PON-CyberSar projects, and by Fondazione Banco di Sardegna. Calculations performed on the Cybersar@UNICA and SLACS-HPC@CASPUR clusters.

%%%%%%%% figure captions
\end{document}